\documentclass[%
 aip,
 amsmath,amssymb,
 reprint,%
]{revtex4-1}

\usepackage{graphicx}
\usepackage{dcolumn}
\usepackage{bm}

\usepackage[utf8]{inputenc}
\usepackage[T1]{fontenc}
\usepackage{mathptmx}
\usepackage{etoolbox}
\usepackage{subfigure}
\usepackage[colorlinks=true, linkcolor=blue, citecolor=blue, urlcolor=blue]{hyperref}

\makeatother
\begin{document}


\title[]{Spatio-temporal Monte Carlo modeling of loading and decay dynamics in an optical dipole trap}

\author{Sankalp Shandilya}
\altaffiliation{Electronic mail: \href{mailto:sankalp@rrcat.gov.in}{sankalp@rrcat.gov.in}}
\affiliation{Laser Physics Applications Division, Raja Ramanna Centre for Advanced Technology, Indore-452013, India}
\author{Gunjan Verma}
\affiliation{UGC-DAE Consortium for Scientific Research, University Campus, Khandwa Road, Indore-452001, India}

\author{Kavish Bhardwaj}
\affiliation{Laser Physics Applications Division, Raja Ramanna Centre for Advanced Technology, Indore-452013, India}

\author{S.P. Ram}
\affiliation{Laser Physics Applications Division, Raja Ramanna Centre for Advanced Technology, Indore-452013, India}

\author{V.\,B. Tiwari}
\affiliation{Laser Physics Applications Division, Raja Ramanna Centre for Advanced Technology, Indore-452013, India}
\affiliation{Homi Bhabha National Institute, Training School Complex, Anushakti Nagar, Mumbai-400094, India}

\author{S.\,R. Mishra}
\affiliation{Homi Bhabha National Institute, Training School Complex, Anushakti Nagar, Mumbai-400094, India}

\date{\today}

\begin{abstract}
 Here, we present our studies on intra-trap dynamics of an optical dipole trap (ODT) loaded from a Magneto-optical trap (MOT) of $^{87}$Rb atoms. A spatio-temporal Monte Carlo (MC) simulation approach has been employed in conjunction with the semi-classical theory based calculations for estimation of various intra-trap two-body loss parameters in the ODT. The temporal MC method based on estimated parameters from spatial MC, gives a close estimation of the experimentally observed atom-number evolution in ODT. We also find that, the decay rate of atoms in an ODT is dominated by momentum-transfer elastic collisions in the absence of MOT beams. However, in presence of MOT beams, the radiative escape process is the dominant two-body loss mechanism in the trap, which surpasses the fine-structure changing and hyperfine changing collisional losses by nearly an order of magnitude. 

\end{abstract}
\maketitle

\section{Introduction}
During the last few decades, considerable progress has been made in understanding various types of cold atom traps. This is because cold atoms serve as an excellent test bed for fundamental science and have immense applications in upcoming quantum technologies, including sensing and quantum information \cite{Zhai_2021}. Optical dipole traps (ODTs) have emerged as vital tools for trapping and manipulating atoms specifically in the domain of quantum computing and Bose-Einstein condensation \cite{article,PhysRevLett.87.010404}. Over the years, there have been significant experimental and theoretical advances in understanding the loading and decay dynamics of an ODT \cite{GRIMM200095}. Early studies have primarily focused on the efficient loading techniques \cite{PhysRevA.62.013406} for transferring a large number of atoms from MOT to ODT with larger beam waists (i.e., beam waists much larger than the optical wavelength). Currently, considerable effort has been devoted to exploring the loading and decay of micro-sized ODT to trap a single atom for quantum information applications \cite{Grünzweig2010,PhysRevA.85.062708}.
\\

To date, the experimentally observed dipole trap loading \cite{GRIMM200095,PhysRevA.47.R4567} involving large atom numbers ($\approx 10^{5}$) has been analyzed by fitting dynamic differential equation governing the trap loading. This fitting has been used to extract parameters such as loading rate, one-body loss rate, and two-body loss rate \cite{PhysRevA.62.013406}. This approach relies on implicit averaging of trap dynamics (assuming constant trap parameters) both spatially and temporally for large atom numbers without explicitly carrying out the trap averaging for relevant physical processes. The effect of near-detuned light on intra-trap loss processes has been carried out extensively on both theoretical and experimental fronts for MOT \cite{PhysRevA.63.043403,PhysRevA.44.4464}. In the case of large atom number in ODT, Kuppens et al. \cite{PhysRevA.62.013406} have briefly discussed the effects of MOT beam on the intra-trap losses in a $^{85}Rb$ dipole trap. However, such analysis requires a much detailed phenomenology of the various intra-trap loss processes. The detailed analysis of intra-trap cold-collisional processes \cite{BUeberholz_2000,JPiilo_2001}have been done in a general context but not specific to the ODT.


In this work, we have developed a spatio-temporal MC approach to study loading and decay dynamics of an ODT, in which intra-trap collisions have been modeled rigorously. Our model begins with the estimation of the intra-trap loss rates using the values obtained from semi-classical models involving spatial MC based trap averaging. We have used the specific parameters for the ODT and MOT (including beam intensity, detuning, and size). The model explicitly addresses the effect of MOT beams on dipole trap losses during loading of a $^{87}Rb$ dipole atom trap. In presence of near-resonant MOT beams, the intra-trap two body loss rate is dominated by light assisted radiative escape process. The other inelastic intra-trap losses are due to fine-structure changing collisions \cite{PhysRevA.44.4464} and bright hyperfine changing collisions \cite{PhysRevA.62.030702} , their contributions are lower by an order of magnitude. In absence of near-detuned MOT beams, the intra-trap two-body losses are dominated by elastic momentum-transfer collisions. The contribution of spin-changing collisions is found to be almost negligible, as no magnetic bias-field or gradient field is present once dipole trap is loaded. The trap depth was calculated from the dynamic polarizability model \cite{PhysRevA.76.052509} applied to the $^{87}Rb$ ground state. From the model, we can extract the two-body collisional parameter which is subsequently used in a temporal MC simulation \cite{PhysRevLett.89.023005} to obtain the complete evolution of the atom number in the trap. Therefore, using a spatio-temporal MC simulation to generate realistic loading and decay data, we developed a robust framework for optimizing real-time atom number behavior in strongly loaded traps. This approach enables us to simulate atom number evolution that closely resembles the experimental behavior. For the first time, we incorporate the Monte Carlo scheme to simulate the behavior of dipole traps with a large number of atoms ($\approx 10^{5}$), in contrast to its previous application to traps with low atom numbers \cite{PhysRevA.47.R4567,PhysRevLett.89.023005}.


\section{Theoretical Framework}
\subsection{Estimating the loss parameters}

In our optical dipole trap setup (beam waist = $12.9 \, \mu m$), for the typical values of atom numbers present at onset ($\approx 10^{4}-10^{5}$), the thermal de Broglie wavelength ($\approx 20 \, nm$) at $65 \,\mu K$ is much smaller than the length scale of packing (micron sized). The system can be approximately treated as a cold Van der Waal gas of Rb atoms where we use a simple kinetic theory along with  collisional cross-sections evaluated from semi-classical approach. \\

In the presence of MOT beams, phenomenon of light assisted collisions (LAC's) \cite{10.1063/1.37325,PhysRevA.85.062708}  comes into picture in the dipole trap loading. These inelastic light induced collisions greatly enhance the overall intra-trap two body loss coefficient in the dipole trap. There are two major loss channels, namely, radiative escape collisions and fine-structure changing collisions. These processes can be modeled by employing Gallagher-Pritchard (GP) \cite{PhysRevLett.63.957,PhysRevA.44.4464} model in the dipole trap.
In this model, we define an atom-pair distribution function dN(R,E) of N $^{87}Rb$ atoms within dipole trap in a spherical shell of Radius R and width dR with pair energy between E and E+dE. The distribution function dN(R,E) is given as,
\begin{equation}
dN(R,E) = \frac{N^2}{2h} \frac{e^{-E/k_B T_{OD}}}{v(R,E)Q} dR \, dE\ .
\label{eq:1}
\end{equation}
Here Q is translational partition function for atoms, v(R,E) is the velocity in this energy range, h is Planck’s constant, $k_{B}$ is the Boltzmann constant and $T_{OD}$ is the temperature of atoms in the dipole trap. The rate of state changing light assisted collisions (LAC) between two atoms can be obtained as
\begin{equation}
K_{LAC} = \int P(R,E)  \sigma(R) \phi  dN(R,E),
\label{eq:2}
\end{equation}
where $K_{LAC}$ is the rate of two body LAC's, $\phi$ is the photon flux, $\sigma(R)$ is the effective cross-section of interaction. The value of $\sigma(R)$ depends on the line shape function which is decided by the effective detuning and the spontaneous emission cross-section. The detuning is affected by dipole-dipole interaction \cite{PhysRevA.50.R906,PhysRevA.85.062708} and AC stark shift due to intense dipole laser beam \cite{Bhardwaj_2021,PhysRevA.87.063408}. P(R,E) \cite{PhysRevA.44.4464,PhysRevLett.63.957} is the dual atom inelastic collision probability when both atoms are within a certain distance ($R_{eff}$) at which the dipole-dipole interaction energy is larger than the total kinetic energy of the system. P(R,E) is sum of the fine structure changing collision \cite{10.1063/1.37325,Dashevskaya1969TheoryOE} probability ($P_{fs}$) and radiative escape probability ($P_{re}$). For radiative escape, the collision rate is,
\begin{equation}
K_{rad}= \int P_{re}  \sigma(R) \phi  dN(R,E)
\label{eq:3}
\end{equation}
. We also define a collision coefficient associated with radiative escape $K_{re}= \frac{K_{rad}}{N^{2}}$ which remains almost unchanged for the trap as the atom-number evolves. Since the major loss channel is radiative escape \cite{PhysRevA.62.013406}, we consider its phenomenology in detail. The role of fine structure changing process has been discussed later.\\ 

 The probability $P_{re}$ is, 
 \begin{equation}
P_{re}=\frac{\sinh(\gamma t')}{\sinh(\gamma (t' + t_{re}))}
\label{eq:4}
\end{equation}
 where, $t_{re}$ is time taken for reaching radiative escape region ($R_{re} \leq 80 \, nm$) from the point of excitation. It also depends on time ($t'$) to go from radiative escape point ($R_{re}$) to the beginning of core region of inter-atom potential ($R_{c}\approx 0.45 \, nm$). The $R_{re}$ depends on trap potential which varies with position in trap and the pair relative distance.  The relative position oscillates back and forth between $R_{re}$ and $R_{c}$, before the spontaneous emission ($\gamma \approx 6MHz$) occurs. These time scales are governed by the dipole trap depth, the initial pair kinetic energy and the spatial position of excitation of the pair within dipole trap \cite{PhysRevA.50.R906}.

 In this work, we have accounted for the non-uniform density of atoms in the dipole trap when applying the GP model. We have also considered the AC-stark shift of atomic energy levels\cite{PhysRevA.76.052509} between $5S_{1/2}$ (F=2, $m_{F}=+2$) and $5P_{3/2}$ (F=3, $m_{F} =3$) while applying the GP-model in our simulations. A spatial Monte Carlo (MC) method is integrated in this model by randomly choosing atom pairs at various positions inside the trap as per Gaussian density distributions in axial and longitudinal directions. The effect of varying trap depth and distance between two atoms in an interacting pair is incorporated into $P_{re}$ and v(R,E) calculations. We eventually evaluate $K_{re}$. The evaluation of Eq. \ref{eq:2} has been done over appropriate range of R and E values. The upper cutoff for R is decided based on comparison of dipole-dipole interaction energy with thermal energy. The upper cut-off for E is kept as twice the trap depth. \\

In addition to this, we have modeled the average radiative escape probability ($\overline{P_{re}}$) using spatial MC without explicitly using the pair-distribution function as used in the GP model. The calculation involves sampling of pair energy and distance between pair atoms based on their respective probability distributions of kinetic energy and distance between atoms. The $\overline{P_{re}}$ is obtained by averaging over these values in the distribution after accounting for position-dependent $P_{re}$ within trap as discussed in modified GP model. In these calculations, we have incorporated the varying line shape function ($\epsilon(R)$) for an atom in the trap \cite{PhysRevA.44.4464}. Using this, we have estimated the LAC loss rate due to radiative escape in the trap. In these calculations, we have assumed that cloud temperature is constant for simplicity. We also assume that the volume of the cloud ($V_{eff}$) remains unchanged. The distribution of pair kinetic energy (f(E) dE) of two $^{87} Rb$ atoms plays a key role in explicitly determining the probability of radiative escape channel and is given by a gamma distribution 
\begin{equation}
f(E) dE = \frac{1}{(6(k_{B}T_{OD})^{3})}E^{2} exp(-\frac{E}{k_{B}T_{OD}}).
\label{eq:5}
\end{equation}
The distribution of pair distances (f(r)dr) within the interaction sphere (r$\leq R_{eff}$) can be modeled as 
\begin{equation}
f(r) dr= \frac{3r^{2}}{R_{eff}^{3}} dr,
\label{eq:6}
\end{equation}
where $R_{eff}$ is effective range over which the dipole-dipole interactions play role. In case of $^{87}$Rb atoms this range is $\approx 220 \, nm $ at temperature $\approx 65 \mu K$. The effective range is the characteristic scale at which the average kinetic energy is comparable to dipole-dipole interaction potential. Finally, the collisional coefficient for two body radiative escape can be estimated as $ K_{re}= \overline{P_{re}} \pi R_{eff}^{2} \overline{\epsilon(R)} v_{avg} $, where $v_{avg}$ is the average relative speed of atoms in the trap. This approach inherently incorporates the saturation in excitation probability (for large intensities of near-detuned light) of atoms due to available photon flux in the system.\\

The rate of fine-structure changing collisions can be estimated from the available data of loss rate for such collisions \cite{10.1063/1.37325}. The loss rate can be extrapolated for our cloud temperature and the light intensity value, after taking the effective volume into account. This fine-structure changing collision loss rate ($\beta_{fs}$) is found to be an order of magnitude smaller than the RE loss rate. An approximate calculation \cite{PhysRevA.62.030702} for bright ground-state hyperfine changing collision can be made, which also yields a loss rate an order smaller. The spin-changing ($m_{F}$) collisions are also present \cite{PhysRevA.66.053616}, but losses from them are significantly weaker in comparison to radiative escape. This is because spin flip does not affect significantly the dipole trap depth for far-off detuned trap.  The elastic collisions loss rate are also an order of magnitude smaller as compared to radiative escape, as discussed in the next section.\\

\begin{figure}[ht]
\centering\includegraphics[width= 11.2 cm]{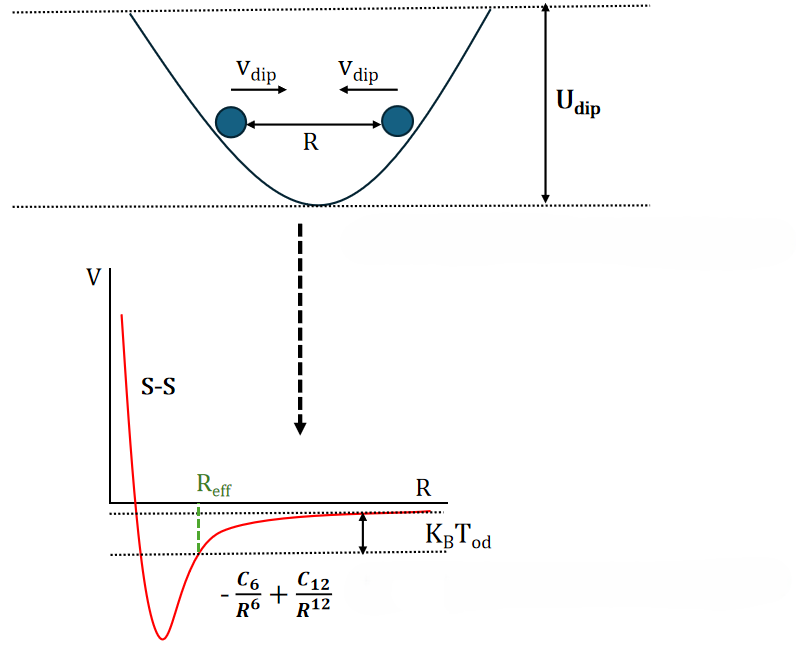}
\caption{A schematic diagram of the interaction between two ground state intra-trap $^{87}Rb$ atoms inside the dipole trap.}
\label{fig:1}
\end{figure}
In the absence of near resonant MOT beams, the intra-trap elastic evaporative collisions induce momentum transfer with energy of an atom sufficient to get kicked out of the dipole trap. They only lead to one atom loss from the trap. In our semiclassical model (Fig. \ref{fig:1}), we incorporate quantum scattering due to low-energy s-wave collisions which are energy independent to approximate the effective cross-section of interaction ($\sigma_{el}$) \cite{PhysRevLett.88.093201}. The two body collision ($R_{el}$) can be approximated as,
\begin{equation}
R_{el}= \frac{\sigma_{el}v_{r}}{V_{eff}} N^{2} ,
\label{eq:7}
\end{equation}

where $\sigma_{el}= 8 \pi a_{s}^{2}$ and $a_{s}$ ($\approx 5.3 \, nm$) is the low energy s-wave scattering length \cite{PhysRevLett.88.093201} for $^{87}Rb$ atoms, $v_{r}$ is average relative velocity of dipole trap atoms in the direction of approach and N is the number of atoms in dipole trap. For our system, the trap frequency in radial direction is $\approx 10$ kHz and in z-direction is $\approx 100$ Hz. Both these values are greater than estimated range of single-atom elastic collision rate ($\frac{N \sigma_{el} v_{r}}{V_{eff}}$). Thus the cold atom cloud system in consideration is in non-hydrodynamic regime and the collision rate scales as $N^{2}$. The effective volume $V_{eff}$ is given as,   
\begin{equation}
V_{eff}= (2\pi)^{3/2} a_{r}^{2} a_{z},    
\label{eq:8}
\end{equation}
where $a_{r} \approx 6 \mu m$ is the extent of the cloud size in transverse direction and $a_{z} \approx 436 \mu m$ is the corresponding extent of the cloud in longitudinal direction of the dipole beam propagation. Similar to the LAC case we define an intra-trap elastic collision coefficient as $K_{el}= \frac{R_{el}}{N^{2}}$. Finally, the two-body loss coefficient can be estimated accounting for the fraction of loss causing evaporative collisions. This is given as,
\begin{equation}
\beta_{dec}= K_{el}\frac{2}{\sqrt{\pi}} \eta e^{-\eta}
\label{eq:9}
\end{equation}
where, $\eta = \frac{U_{dip}(r,z)}{k_{B}T_{OD}}$ dictates the probability of loss events. Here, $U_{dip}(r,z)$ is the position dependent dipole trap potential. The Eq. \ref{eq:9} is evaluated by using spatial MC accounting for position dependent density and trap potential as discussed before for radiative escape. There is only a fraction of elastic collision events \cite{KETTERLE1996181} within the effective range leading to loss of an atom. \\

Further, in absence of MOT beams, where we study decay dynamics in dipole trap, the contribution of dark ground-state hyperfine changing collisions \cite{PhysRevA.62.030702} to the overall loss rate can be neglected. This is because most atoms in our system are in the dark (F=1) ground state as the re-pumper beams are turned off slightly before the MOT beams. The hyperfine energy scale ($\approx 350 \mu K$) is much higher than the available energy scale ($\approx 65 \mu K$) of the trapped atoms in the ODT. Thus, this collision channel is effectively suppressed. The spin-changing collisions are also suppressed as there is no Zeeman splitting because magnetic fields are turned off during decay of dipole trap. Only the far-detuned dipole beam is present so no light assisted losses are possible.

\subsection{Temporal evolution}
\begin{figure}[ht]
\centering\includegraphics[width= 8.5 cm]{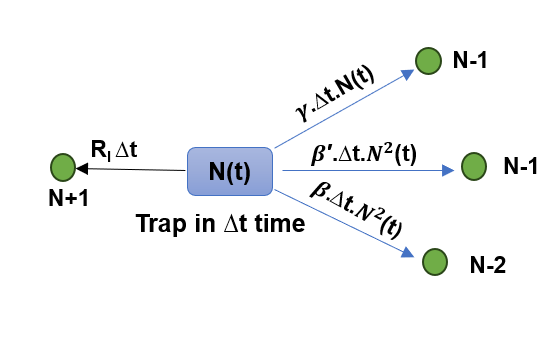}
\caption{A schematic diagram of the possible processes in a dipole trap in a very small time window ($\approx$ 20 ns) and their associated probabilities.}
\label{fig:2}
\end{figure}

The simulation starts with an empty trap in loading studies and probabilities of events are then updated with time as the trap gets filled. For the decay analysis, the initial condition is determined via spline interpolation of the actual data and the probabilities are updated similarly. The role of the physical processes involved in loading and loss \cite{PhysRevLett.89.023005,Wang:23,Fung_2015} of atoms into and from the trap, is accounted for by considering the relative probabilities of the possible events in the dipole trap in a preset time scale ($\Delta t$). A number is chosen randomly from the uniform distribution (between 0 and 1) and compared with the relative probabilities of the possible events, and based on the comparison, one of the events is chosen. These events include probability of loading of a single atom (a(t)) , background atom collisions loss (b(t)) probability, two body intra-trap single atom loss (c(t)) probability, two body intra-trap dual atom loss (d(t)) with probability, and probability of no event at all. These four event probabilities (Fig. \ref{fig:2}) are calculated by choosing the value of $\Delta t$ and using the trap loading and collision parameters. The number of atoms in the trap is then updated. The data generated from MC run is compared with the experimental data and the final parameters are obtained by minimizing the point by point absolute error from the actual data.
\begin{equation}
a(t)= R_{l}\Delta t    
\label{eq:10}
\end{equation}
\begin{equation}
b(t)= \gamma N(t)\Delta t  
\label{eq:11}
\end{equation}
\begin{equation}
c(t)= \beta^{'} N^{2}(t) \Delta t  
\label{eq:12}
\end{equation}
\begin{equation}
d(t)= \beta N^{2}(t) \Delta t  
\label{eq:13}
\end{equation}
\\

Here N(t) is the number of atoms in the trap at time t, $R_{l}$ is the atom loading rate in the dipole trap from MOT, $\gamma$ is the background collision loss coefficient, $\beta$ is the two body intra-trap collisional loss coefficient for processes leading to two atom losses (inelastic state changing collisions), $\beta^{'}$ is the two body loss coefficient for intra-trap elastic losses which only lead to single atom loss. The estimation of these loss parameters has already been discussed in the prior subsection using semi-classical models. The value of loading rate \cite{PhysRevA.62.013406} $R_{l}= \frac{1}{4}n_{MOT}v_{MOT}A_{ODT}P_{trap}$ is estimated using kinetic theory approach. Loading rate is directly proportional to the incident flux of MOT atoms ($n_{MOT}v_{MOT}$) on the cylindrical surface of the dipole trap ($A_{ODT}$) and trapping probability ($P_{trap}$) which depends on the dipole trap depth and MOT cloud temperature. The value of $\gamma = n_{BG}v_{BG}\sigma_{Rb-X}$ is also estimated from kinetic theory approach \cite{PhysRevA.106.052812,Steane:92} which depends on background hot atom flux ($n_{BG}v_{BG}$) as well as the interaction cross-section ($\sigma_{Rb-X}$) for elastic momentum-transfer collisions of Rb with all atoms in background\cite{Steane:92}. \\

Our methodology is applicable to understand the dipole trap loading and decay over a wide range of operational parameters including a range of beam waist size, a range of trapping potential, range of MOT atom cloud temperature, etc. Similarly, the methodology is applicable during the decay analysis.

\section{Experimental}
\begin{figure}[h]
\centering\includegraphics[width=8 cm]{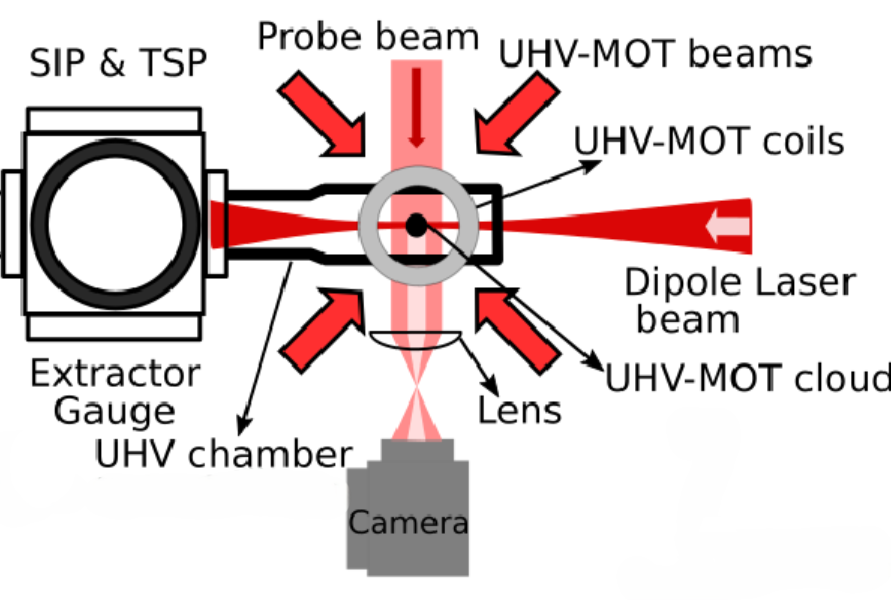}
\caption{The experimental setup of single MOT chamber loaded single beam dipole trap}
\label{fig:3}
\end{figure}

\begin{figure}[h]
\centering\includegraphics[width=8 cm]{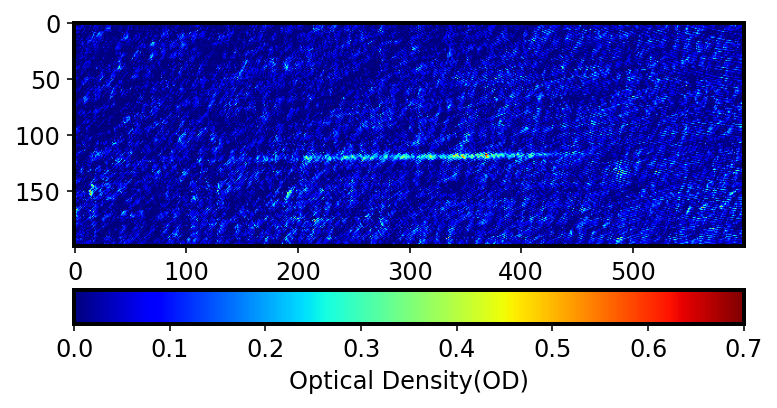}
\caption{The experimental optical density image of the trapped atom cloud obtained by absorption imaging}
\label{fig:4}
\end{figure}

We have experimentally studied the loading and decay dynamics of a dipole trap of beam waist 12.9 $\mu m$ ($\frac{1}{e^{2}}$ radius) . Our experiments (Fig. \ref{fig:3}) have been performed using a single-chamber magneto-optical trap (MOT) for laser-cooled $^{87}$Rb atoms \cite{10.1063/5.0106398} which has been used to load a single-beam optical dipole trap (ODT). The setup consists of a glass-cell ultra-high vacuum (UHV) chamber (UHV-MOT) at $\sim 1.5 \times 10^{-10}$~Torr. Cooling beams for the MOT are generated using a high power tunable diode laser system (TOPTICA TA-PRO). The cooling laser is detuned by 15 MHz from the $5S_{1/2}, F=2 \rightarrow 5P_{3/2}, F'=3$ transition of $^{87}$Rb atom. The output of the laser system is expanded to form the UHV-MOT (six-beam configuration). The repumper laser beams are generated using an extended cavity diode laser system (TOPTICA DLC-PRO) which is stabilized to the transition the $F=1 \rightarrow F'=2$ of $^{87}$Rb atom. The quadrupole magnetic fields is provided by two coil pairs, and Rb vapor is supplied by resistively heated Rb-dispensers. With proper laser locking, magnetic field tuning, and beam alignment, we routinely load $\sim 5\times 10^7$ atoms into the UHV-MOT at a temperature of $\sim$50~$\mu$K after implementing molasses stage. 
The dipole trap is formed by using the output of a master oscillator power amplifier (MOPA) system (TOPTICA TA PRO) operating at wavelength 811~nm. The laser beam was focused onto the UHV-MOT cloud using a 200~mm focal length lens mounted on a 3-axis translation stage. The trap is loaded from the MOT in a duration of about 100 ms. The optical dipole trap (ODT) was loaded by switching-on the dipole laser beam on MOT cloud. The trapped cloud in ODT was imaged by absorption probe imaging technique. Using an appropriate image processing software, the absorption images were converted into optical density (OD) images. A fine straight line at the center of the image of atom cloud was observed at dipole laser power of $\approx$ 130 mW (corresponding to a trap depth of $\approx 77 \mu K$). This showed the trapping of cold atoms in ODT as seen in Fig. \ref{fig:4}. The lifetime study has been performed for 12.9 $\mu m$ trap after turning off the MOT beams. The single MOT beam intensity at the position of dipole trap center is around 17 $mW/cm^{2}$.

\section{Results and Discussion}
\subsection{Simulation results for loss parameters}
\begin{figure}[h]
\centering\includegraphics[width=8 cm]{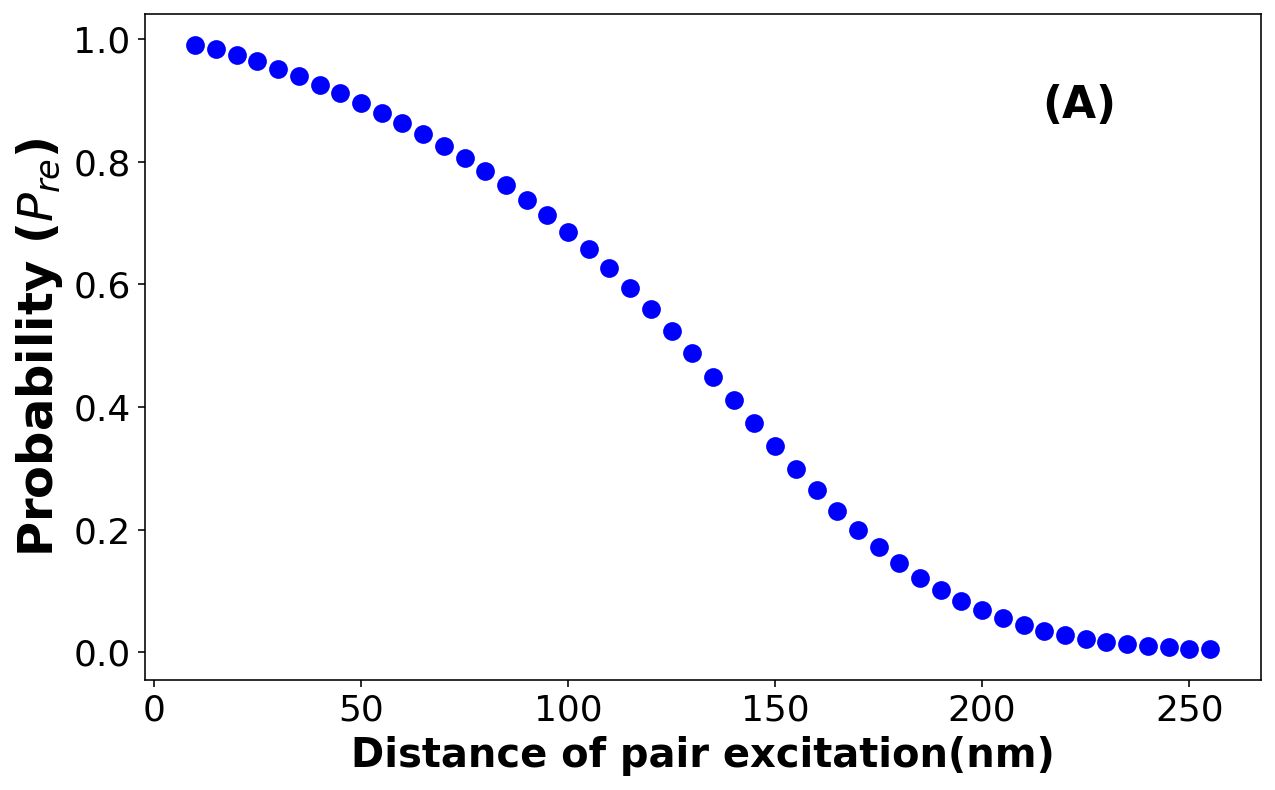}
\centering\includegraphics[width=8 cm]{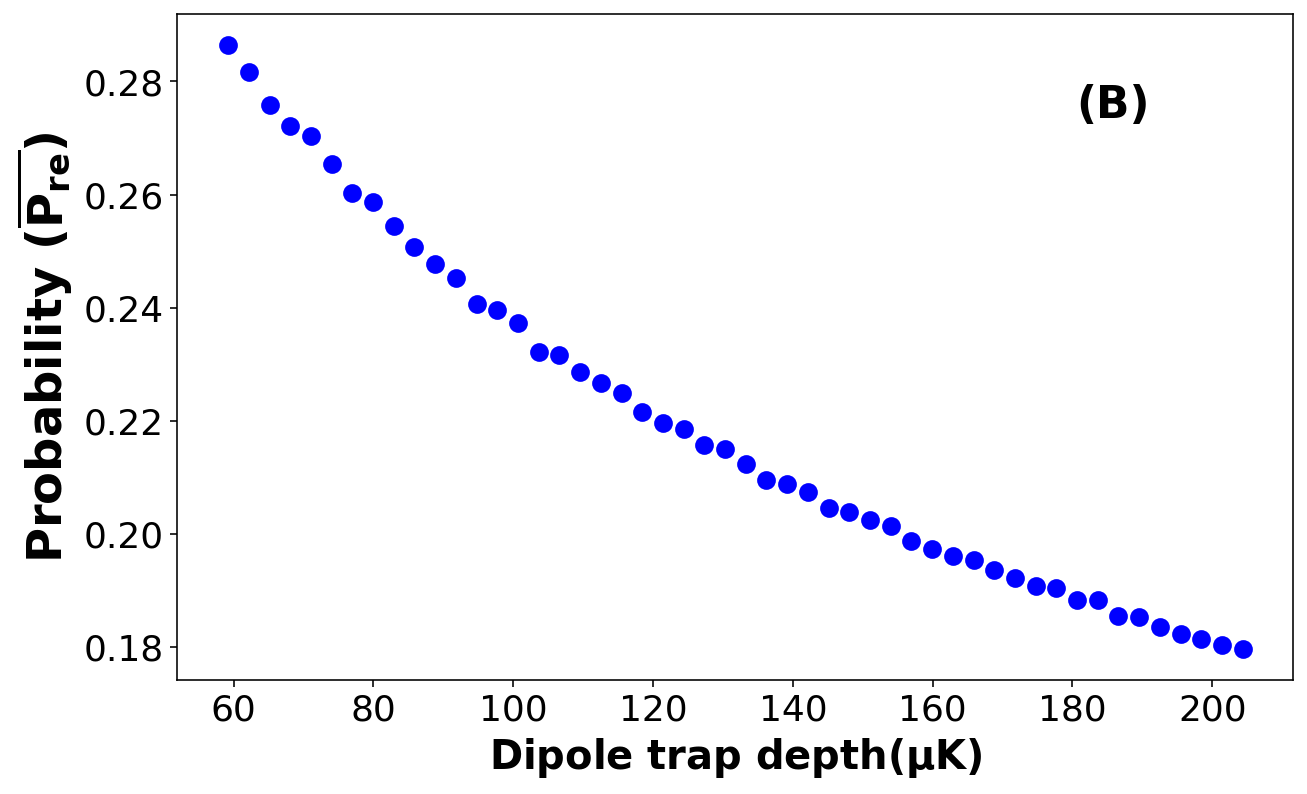}
\caption{\textbf{(A)} The probability of radiative escape collisions with the distance of pair excitation for 12.9 micron trap. \textbf{(B)} Averaged probability of radiative escape collision as function of dipole trap depth. The probability is averaged over all excitation lengths. The sampling is done using distributions discussed in section II. }
\label{fig:5}
\end{figure}

The probability of radiative escape was calculated using the model developed in section II . In all these calculations, cloud volume $V_{eff} \approx 2.6 \times 10^{-7} cm^{3}$ estimated based on temperature obtained from OD images and Eq. \ref{eq:8} was taken as constant. We also assumed a temperature roughly proportional to the trap depth \cite{PhysRevA.62.013406}. The probability for radiative escape ($P_{re}$) after excitation at a certain pair distance (Fig. \ref{fig:5} A) within the atom cloud is found . The value is averaged over all pair kinetic energies sampled from Maxwell-Boltzmann distribution of pair kinetic energy (Eq. \ref{eq:5}) and trap positions sampled from normal gaussian distributions taking density variation into account. The variation of $P_{re}$ is also shown with varying dipole trap depth (Fig.  \ref{fig:5} B) and averaged over pair distance and pair energy distribution (Eq. \ref{eq:5}) .
\begin{figure}[h]
\centering\includegraphics[width=8 cm]{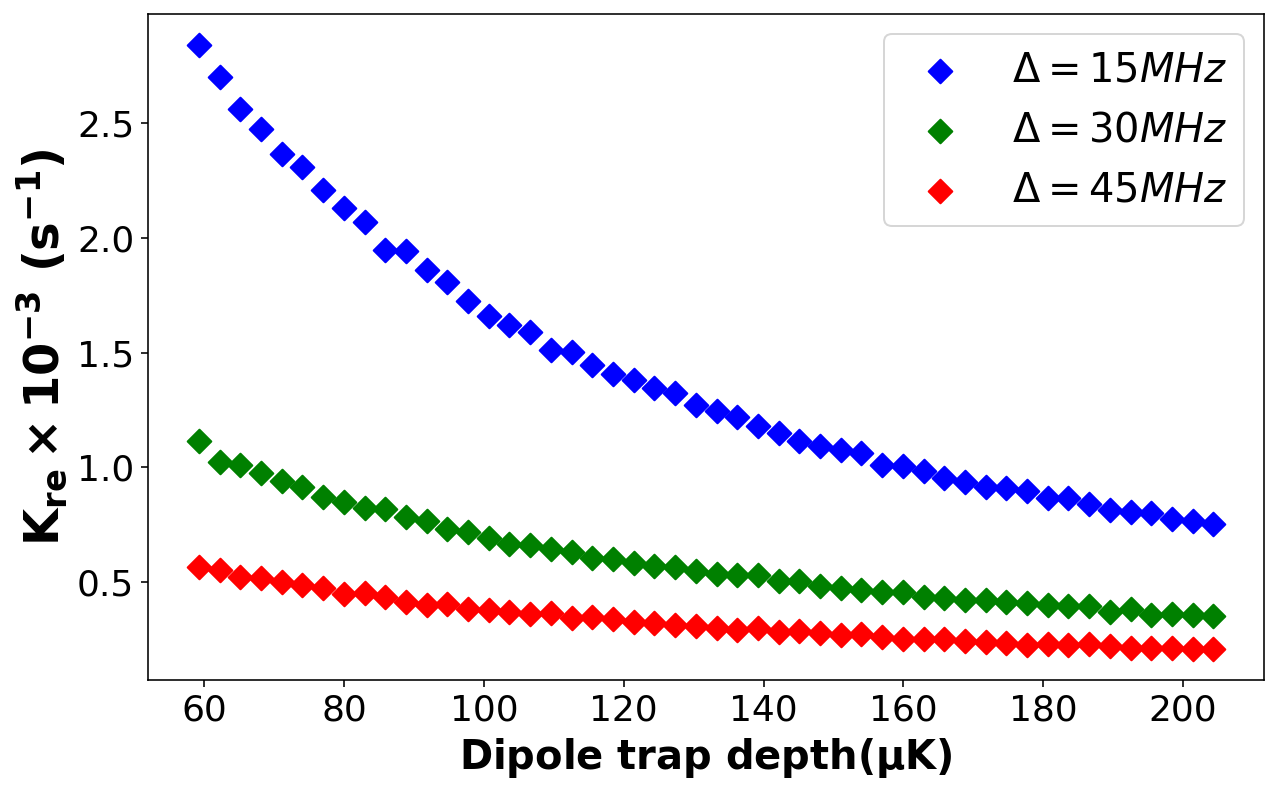}
\caption{The coefficient of radiative escape collisions using modified GP model for 12.9 $\mu$m trap with the dipole trap depth and three different values of red detuning of near-detuned MOT beams is shown. The points are generated via spatial MC simulation.}
\label{fig:6}
\end{figure}
\begin{figure}[h]
\centering\includegraphics[width=8 cm]{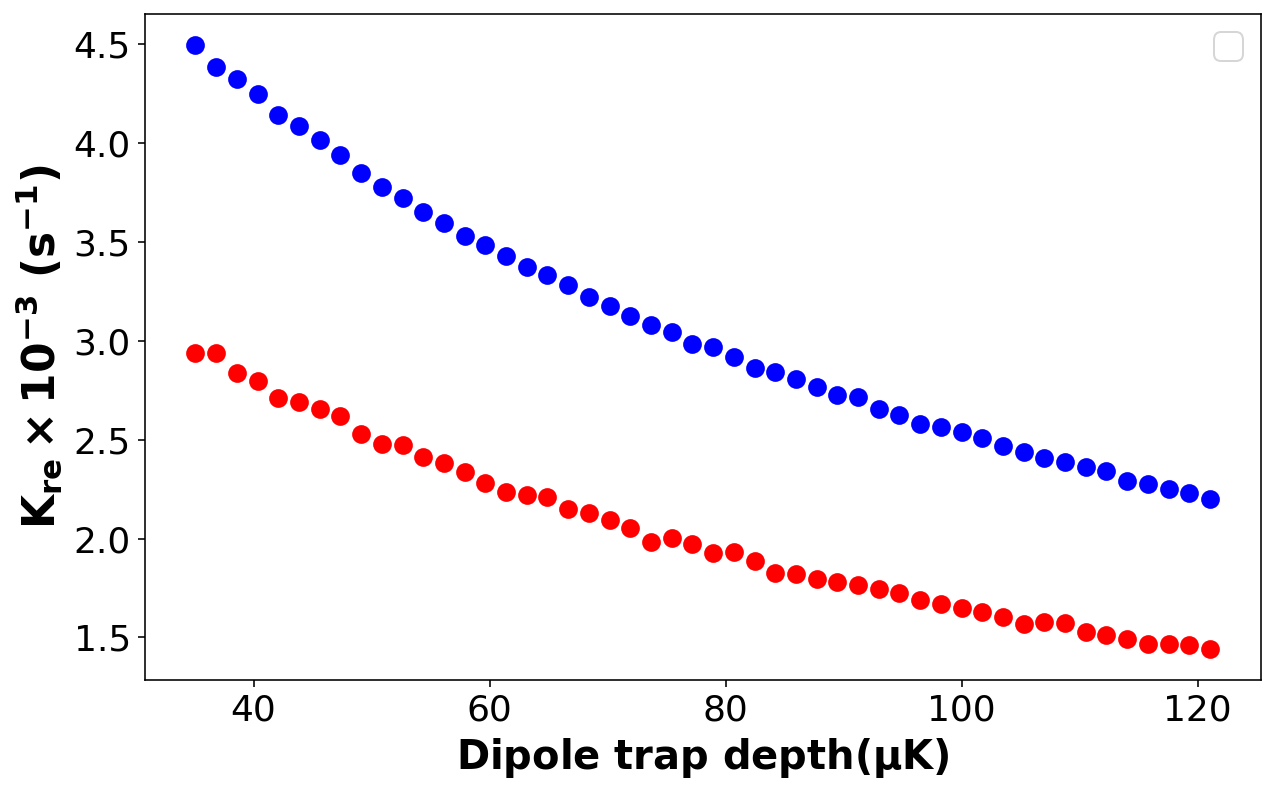}
\caption{The coefficient of radiative escape collisions with the dipole trap depth for 12.9 micron beam waist is shown. The blue circles are using average radiative escape probability model and the red circles are using modified GP model. The points are generated via spatial MC simulation.}
\label{fig:7}
\end{figure}
We have also calculated the $K_{re}$ (shown in Fig. \ref{fig:6}) using GP model (Eq. \ref{eq:3}) integrated with spatial MC with varying trap depths (due to power as beam waist is kept constant). We see that as expected the collision rate decreases with increasing AC stark shift (or power) due to larger net detuning observed for excitation and reduced net absorption cross section. Similarly, we also see a decrease in the calculated collision rates with increase in detuning of MOT beams due to decrease in net absorption cross-section. We have also compared the calculated results (Fig. \ref{fig:7}) for both the models of radiative escape discussed with varying power. The semi-classical model without pair-distribution function also shows a decrease in collision rate as power is increased. The inelastic loss coefficients are simply twice the collision coefficients as each collision effectively results in both atoms being lost.\\

\begin{figure}[h]
\centering\includegraphics[width=8 cm]{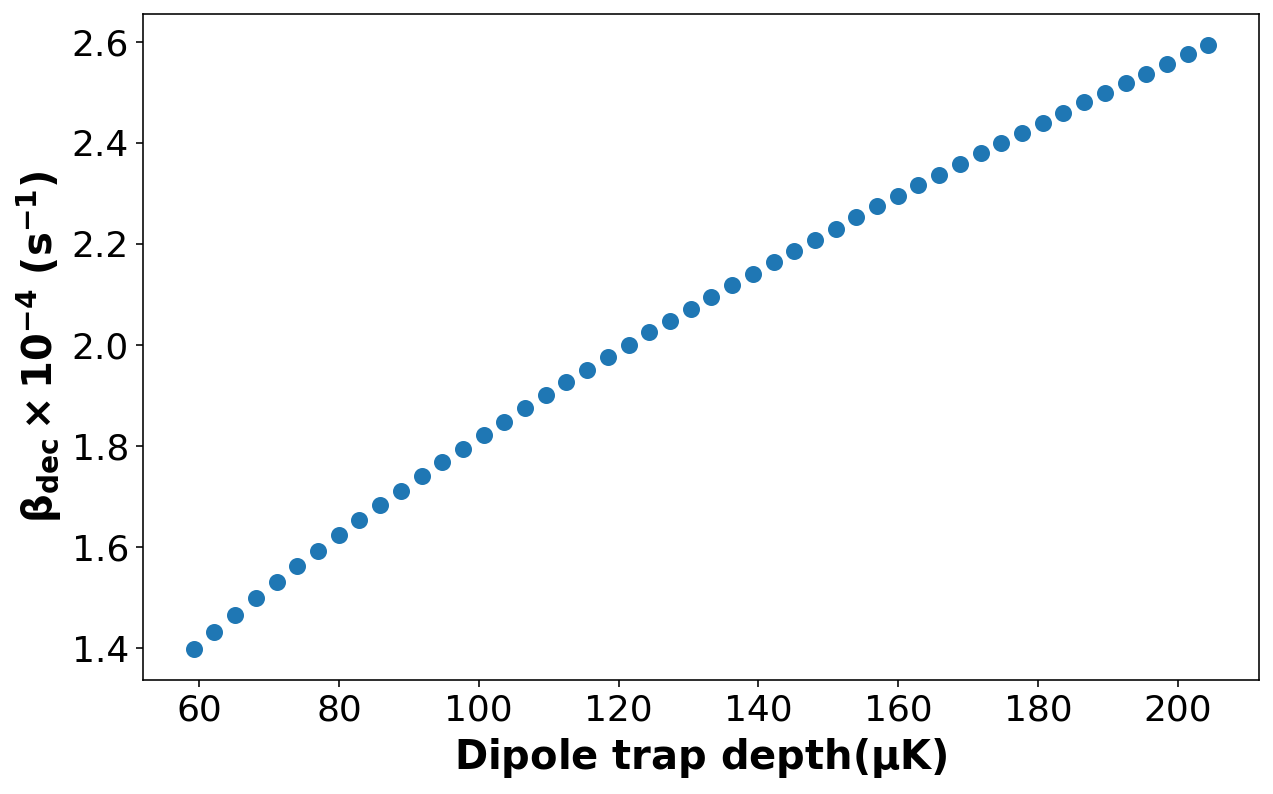}
\caption{The coefficient of elastic evaporative collisions with the dipole trap depth for 12.9 micron beam waist is shown. The points are generated via spatial MC simulation.}
\label{fig:8}
\end{figure}

In the absence of near-detuned MOT beams, we have simulated the elastic evaporative loss coefficient ($\beta_{dec}$) variation with the trap depth (Fig. \ref{fig:8}). We assume a constant cloud volume ($V_{eff}$) for the trap depths considered and a temperature roughly proportional to trap depth as mentioned above for radiative escape. The loss increases as trap becomes tighter due to these elastic collisions. Each such collision only leads to a single atom being lost from the trap.

\subsection{Optical dipole trap loading and decay}

The value of the estimated loading rate for the MOT parameters of density ($\approx \frac{10^{10}}{cm^{3}}$) and temperature ($\approx 50 \mu K$) is around $ R_{l} \approx 10^{6}$ atoms/s. The estimated value for $\gamma=0.25 s^{-1}$ is known from the loss rate of MOT atoms. This serves as a decent approximation to initiate the temporal MC as the $\gamma$ is mainly dependent on background same for the traps. A more accurate estimate is not needed as the dynamics is dominated mainly by intra-trap losses. The estimated values of two body loss coefficients for fine-structure changing collisions ($\beta_{fs}\approx 2\times10^{-4} s^{-1}$), bright ground state hyperfine changing collisions ($\beta_{hf}\approx 1.9\times10^{-4} s^{-1}$) and the radiative escape collisions ($\beta_{re} \approx 4.6\times10^{-3} s^{-1}$) are calculated. This gives a collective two body loss coefficient $\beta_{LAC}\approx 5\times10^{-3} s^{-1}$. However, if we use the estimate of two body loss coefficient ($\approx 8\times10^{-3} s^{-1}$) obtained using average radiative escape probability model we get collective two body loss coefficient $\beta_{LAC}\approx 8.4\times10^{-3} s^{-1}$.  \\

\begin{figure}[h]
\centering\includegraphics[width=8.5 cm]{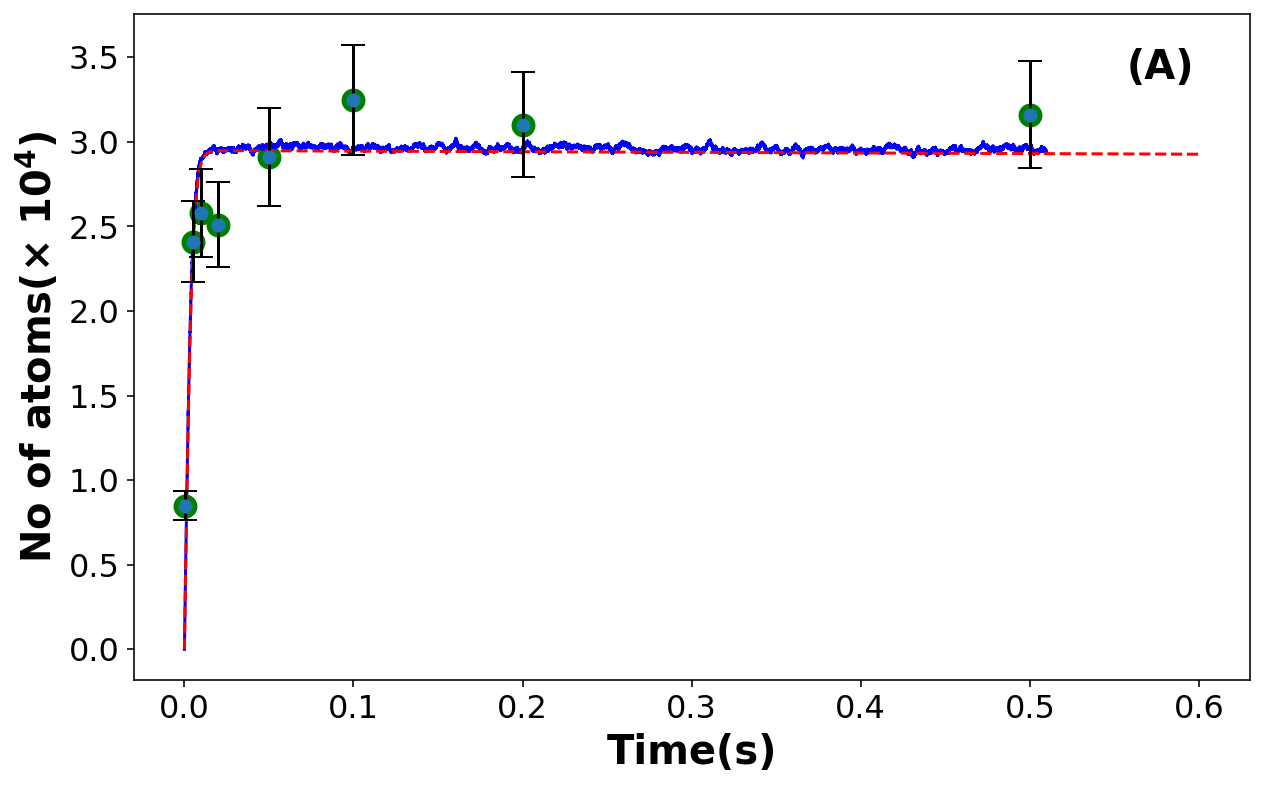}
\centering\includegraphics[width=8.5 cm]{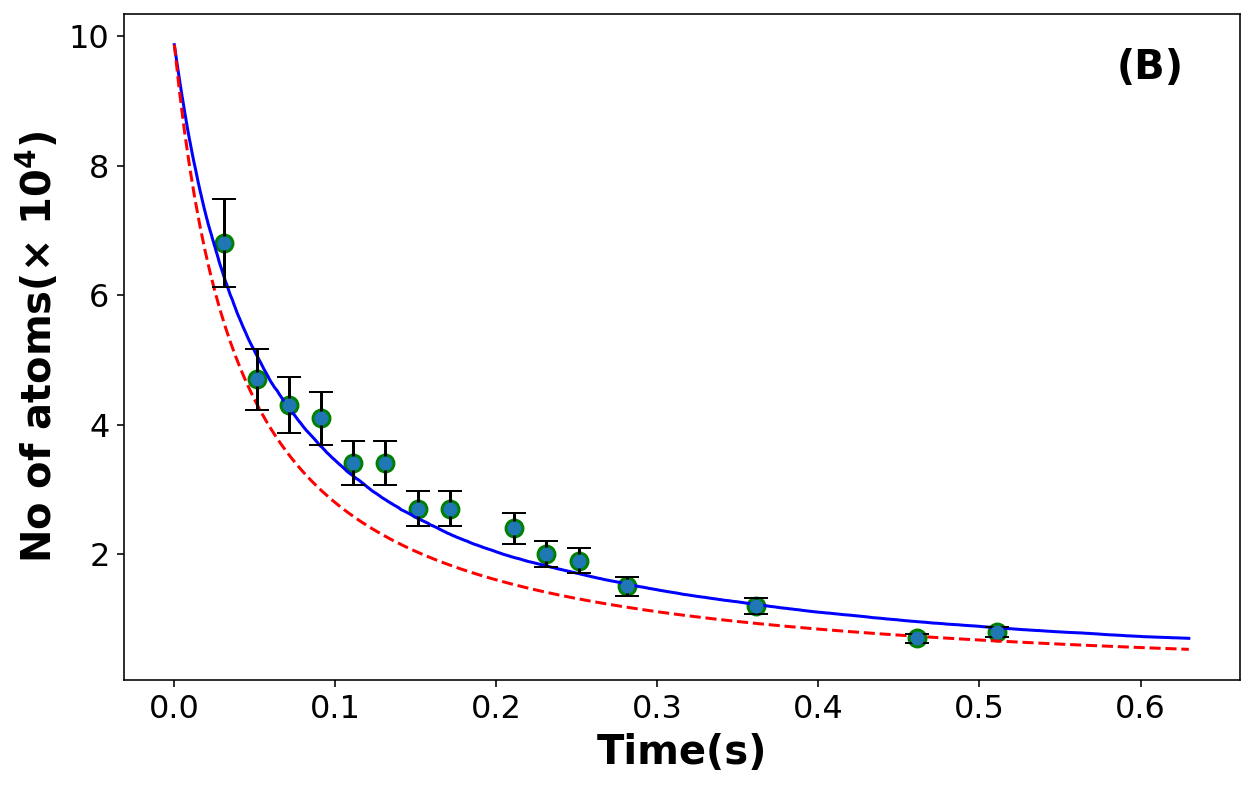}
\caption{\textbf{(A).} The Monte Carlo interpolation and differential equation fit to the experimental dipole trap loading data for 12.9 $\mu$m is shown. The continuous blue line is the MC fit and dashed red curve is the differential equation fit to the data (represented by green circles). \textbf{(B).} The MC interpolation and differential equation fit to the experimental dipole trap decay data for 12.9 $\mu$m trap is shown. The indicatives are same as loading curve}
\label{fig:9}
\end{figure}

The estimated parameters are used to initiate the MC data generation. The value of the final parameters obtained from our MC method (Fig . \ref{fig:9}A) applied on the experimental data are $R_{load} = 6.7\times10^{6} s^{-1}$, $\gamma =0.2 s^{-1}$ and $\beta_{MC}  =7.6\times10^{-3} s^{-1}$. The value of the estimated two body loss coefficient is in good agreement with the experimentally observed value. The differential equation model \cite{PhysRevA.44.4464} (Fig. \ref{fig:9}A) with the inclusion of loading term has been used as well to fit the data as well and the loss coefficient obtained is $\beta_{DE}= (7.4 \pm 0.7) \times 10^{-3} s^{-1}$. In all the earlier studies concerning cold $Rb^{87}$ atoms it has been observed that fine-structure changing collisions is a dominant loss channel as far as MOT is concerned. This observation of tremendous enhancement of radiative escape channel can be attributed to the very small trap depth of our dipole trap compared to MOT trap depths.
 \\

For decay dynamics, the estimated value for $\gamma=0.25 s^{-1}$ and that of $\beta^{'}_{decay}=1.6\times10^{-4} s^{-1}$ is obtained by the model of elastic evaporative loss discussed for a cold Van der Waals gas. The values given by the experimental data via Monte Carlo fitting (Fig. \ref{fig:9} B) and error minimization are $\gamma = (0.4\pm0.15)s^{-1}$, $\beta=(1.8\pm0.4)\times10^{-4} s^{-1}$. The fitting via differential equation has been shown in same plot, suggesting a better fit via MC technique. The value of $\beta$ obtained from differential equation fitting is $(2.4\pm0.2)\times10^{-4} s^{-1}$. The semiclassical model seems to adequately capture most of the features of the collisions.\\

We have compared the MC evolution to the values obtained by the conventional differential equation (DE) fitting by comparing the point by point absolute error for both fitting schemes. In trap decay, the values are 10$\%$ and $18 \%$ respectively. For the case of loading, the DE fitting gives error of $19\%$ with MC being slightly better at $18\%$ . This is because for loading case the parameter which dominates the fitting during rapid rise is $R_{l}$, which is estimated easily by the optimization algorithm due to the initial linear increase in atom number independent of existing trap number. But, for the decay case the $\beta$ term of two body loss (has non-linearity) dominates the fitting (atom number dependent loss) which has to be optimized by the DE method and hence the difference is clear. Our overall MC  technique based on estimation of parameters gives better results in terms of fitting accuracies as compared to the conventional differential equation fitting. This establishes that not only is the MC approach applicable for higher number of atoms in the dipole trap, but is slightly better in terms of resemblance to the actual experiment.

\section{Conclusion}

We have used Monte Carlo method to estimate various intra-trap loss processes inside the optical dipole trap loaded from a MOT. The loading and decay dynamics have been simulated using the trap loading and loss-rates obtained from semiclassical theory. Spatial MC was used to obtain the intra-trap loss rates in presence and absence of MOT beams. From the model, we find that radiative escape is the dominant loss process in presence of near detuned MOT beams. In the absence of MOT beams, elastic evaporative collisions dominate the trap loss. The atom number evolution obtained from the temporal MC scheme explains the experimentally observed data to a fair extent.

\section{Acknowledgment}
We are thankful for the scientific discussions with Vivek Singh, Surendra Singh, Subhajit Supakar and control electronics support from Ayukt Pathak and Shradha Tiwari in this work. \\

\noindent The authors declare no conflict of interest.

\bibliography{reference}

\end{document}